\documentclass[aps,prl,twocolumn,groupedaddress,showpacs]{revtex4-1}

\usepackage[pdftex]{graphicx}
\usepackage{epstopdf}
\usepackage{color}
\usepackage[normalem]{ulem}
\begin{document}

\title{Control of gradient-driven instabilities using shear Alfv\'{e}n
  beat waves}
\author{D.W. Auerbach}
\author{T.A.  Carter}
\author{S.  Vincena}
\author{P. Popovich}
\affiliation{Department of Physics and Astronomy, University of
  California, Los Angeles, CA 90095-1547}


\date{\today}

\begin{abstract}
A new technique for manipulation and control of gradient-driven
instabilities through nonlinear interaction
with Alfv\'{e}n waves in a laboratory plasma is presented. A narrow,
field-aligned density depletion is created in the Large Plasma Device
(LAPD), resulting in coherent, unstable fluctuations on the periphery
of the depletion.  Two independent shear Alfv\'en waves are launched
along the depletion at separate frequencies, creating a nonlinear
beat-wave response at or near the frequency of the original
instability. When the beat-wave has sufficient amplitude, the original
unstable mode is suppressed, leaving only the beat-wave response, generally at lower amplitude.
\end{abstract}

\maketitle

In magnetized plasmas in the laboratory, the loss of heat, particles,
and momentum across the confining magnetic field is predominantly
caused by turbulent transport associated with pressure-gradient-driven
instabilities~\cite{Horton:1999p40, Tynan:2009p1232}.  
Controlling these instabilities to mitigate the
transport they produce is therefore highly desirable for magnetic
confinement fusion devices such as tokamaks.  
Direct control of gradient-driven instabilities has been achieved using arrays of
electrodes or external saddle coils to induce fluctuating parallel
currents in a cylindrical plasma device~\cite{Schroder:2001p17,
  Schroder:2004p666}.  In these experiments, when the excited electric
fields matched the frequency and wavenumber of a drift-wave mode in
the device the observed turbulent spectrum collapsed onto the single
coherent driven mode (synchronization) and turbulent transport is
reduced~\cite{Brandt:2009p1425}.  Extending the transport control
enabled through mode-synchronization to high-temperature
fusion devices is highly desirable but it would not be possible using
material electrodes.  It is also not clear whether saddle coils would
be effective at driving currents for synchronization in a
high-performance device.

Turbulence control with core plasma access might be achieved using
externally-launched radiofrequency (RF) waves.  High-power RF is
expected to be an important tool for heating and current drive on
next-step fusion experiments such as ITER~\cite{Shimada:2007p1419} and
could be available for application to turbulence and transport
control.  Creation of flow shear for turbulence
suppression~\cite{Terry:2000p1409} using RF waves has been
theorized~\cite{Myra:2002p1414} and recently demonstrated in the
Alcator C-Mod tokamak~\cite{InceCushman:2009p1418, Lin:2009p1408}.
Direct modification of gradient-driven instabilities is also possible
through interaction with an RF field.  Control of drift-wave
fluctuations was demonstrated in Q-machine experiments using lower
hybrid waves to affect plasma properties and the drift-wave dispersion
to suppress growth~\cite{Wolf:1980p462, Wong:1978p515, Gore:1978p532,
  Liu:1980p361}.  The modification of other transport
  related instabilities, such as the ion temperature gradient (ITG)
  instability, in the presence of RF driven fast magnetosonic waves
  has also been suggested and investigated
  theoretically~\cite{Chiu:1989p340}.  

This Letter reports on novel laboratory experiments using RF waves, in
particular shear Alfv\'{e}n waves (SAWs), to modify
gradient-driven instabilities.  Two co-propagating SAWs with slightly
differing frequency are launched along a narrow field-aligned density
depletion on which coherent (dominant single-frequency) unstable
fluctuations are observed.  Modification and control of the
instability is seen when the nonlinear beat response exceeds an
amplitude threshold and the frequency of the beat wave is near the
frequency of the gradient-driven instability.  In this case, the
original unstable mode is reduced in amplitude or suppressed
completely, leaving only the beat-driven response, generally at lower
amplitude.  A reduction in broadband fluctuations is also observed
accompanying the suppression of the primary mode.  This control is
observed for a finite range of beat wave frequencies around the
unstable mode frequency; this range widens with increased beat wave
amplitude.

This research was carried out on the upgraded Large Plasma Device
(LAPD) at the Basic Plasma Science Facility (BAPSF) at
UCLA~\cite{Gekelman:2010}.  The LAPD creates a 17.5m long, $\sim$0.6m
diameter cylindrical magnetized plasma.  The plasma is created at 1 Hz
using an indirectly-heated barium oxide coated cathode discharge
(discharge length is typically 10-20ms).  Bulk parameters for the
helium plasmas used in these experiments are: $n_e \sim
10^{12}$cm$^{-3}$, $T_e \sim 6$eV, and $B=800$G.  Ion temperature in
LAPD is typically $T_i \lesssim 1$eV~\cite{Palmer:2005p1420}, although
it was not measured directly in these experiments. 
Under these conditions, the electron thermal speed is greater than the Alfv\'en wave speed (the Kinetic Alfv\'en Wave regime \cite{Morales:1997p11}).

In these experiments, temperature, density and floating potential are
determined using a triple Langmuir probe.  The temperature and density
measurements have been quantitatively verified using microwave
interferometry and swept Langmuir probe measurements. Magnetic field
fluctuations were measured using magnetic pickup coils, differentially wound to
eliminate electrostatic pickup.  The probes are mounted on motorized,
computer controlled probe drives.  Two-dimensional average profiles
are acquired over many repeatable discharges by moving a single probe
shot-to-shot.  Two-dimensional cross-correlation functions are
achieved using a second, stationary reference probe separated axially
(along the background field) from the moving probe.

\begin{figure}[tbp]
\begin{center}
\includegraphics{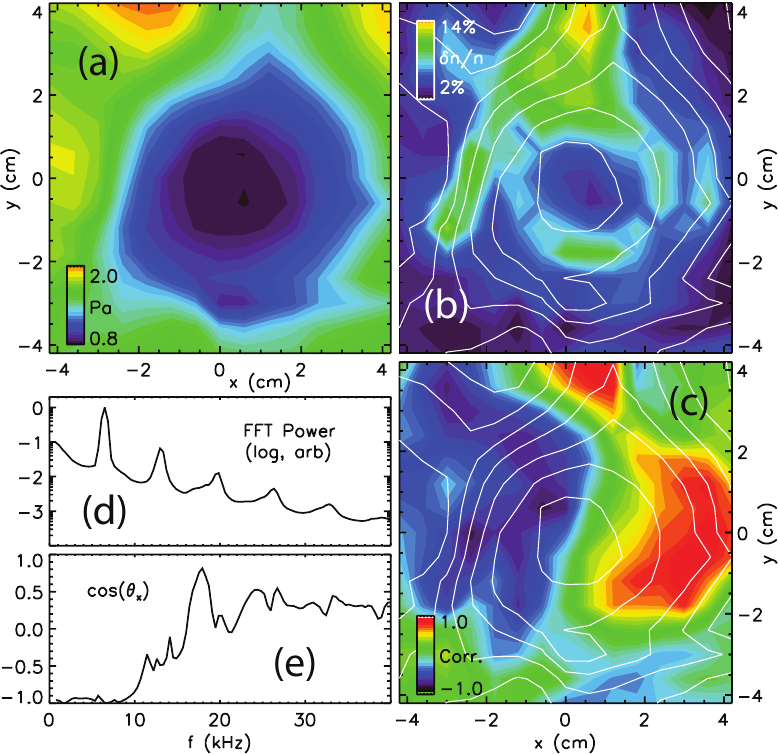}
\caption{[Color]
(a) Contours of measured plasma pressure in the depletion. 
(b) Contours of $I_{\rm sat}$ fluctuation power, with pressure contours overlaid in white.
(c) Cross-correlation showing m=-1 structure of mode.
(d) Spatially averaged $I_{\rm sat}$ FFT spectrum.
(e) Spatially averaged cross-phase ($\theta_x$)between $I_{\rm sat}$ and $V_f$ fluctuations}
\label{fig:density}
\end{center}
\end{figure}

Conditions are established for the instability formation through
creation of a field-aligned axial density depletion embedded in the
center of the LAPD plasma column.  This type of condition has been
extensively investigated in the LAPD, and the presence of
gradient-driven drift-Alfv\'en-wave instabilities in such conditions
is well documented \cite{Maggs:1997p2, Penano:2000p22}.  The
depletion is formed by inserting a small metal blocking disk between
the anode and cathode, selectively blocking the high energy primary
electrons that ionize the neutral gas and leaving a lower temperature
and density striation along the central axis of the machine.  As shown
in Figure~\ref{fig:density}(a) the pressure drops by about 50\% in the
depletion, which is $\sim$6 cm wide.  The measurements shown are of
ion saturation current ($I_{\rm sat} \propto n_e \sqrt{T_e}$) and
pressure $P_e$, with $P_e$ calculated from $T_e$ and $I_{\rm sat}$.
Figure~\ref{fig:density}(b) shows measured $I_{\rm sat}$ fluctuations,
which are localized to the density gradient associated with the
depletion.  Figure~\ref{fig:density}(d) shows the spatially-averaged
fast Fourier transform (FFT) power spectrum of $I_{\rm sat}$,
indicating a narrow-band spectrum with power at $\sim 6$ kHz and
harmonics.  Cross-correlation measurements shown in
Figure~\ref{fig:density}(c) of the spontaneous fluctuations indicate a
$m=-1$ azimuthal mode number for the dominant mode, with propagation
in the electron diamagnetic direction. Many of the measured properties
of the unstable modes (frequency, direction and speed of propagation)
agree with predictions for drift waves.
However, Figure~\ref{fig:density}(e) shows that the measured phase between
density fluctuations and potential fluctuations ($\theta_{x}$) is
approximately $180^\circ$, which is inconsistent with the $ 0^\circ< \theta_{x} < 90^\circ$
predictions and observations of saturated, single mode drift waves \cite{Hendel:1968p1481}.

It is important to note that in these experimental conditions there are two sources of free energy, 
the pressure gradient and an imposed potential gradient from the blocking disk, 
resulting in two separate flows from the $E\times B$ drift and the $\nabla \times P$ drift. 
Thus there are two possible instabilities, the drift-wave instability, 
resulting in drift-waves with axial mode number $n_{\parallel} \ge 0.5$, 
and the Kelvin-Helmholtz instability, resulting in ``flute-like'' modes with $n_{\parallel}=0$.
Numerical eigenmode solutions to the linearized Braginskii two-fluid equations \cite{Popovich:2010p1427} were
performed, using experimentally measured density, temperature, and potential profiles. 
These solutions indicate that both types of modes are linearly unstable and
have real frequency approximately equal to the measured frequency, leaving open the possibility that the saturated single unstable mode observed might be of either type of instability or a hybrid mode driven by both sources of free energy.

A two-strap ``picture-frame'' style
  antenna~\cite{Zhang:2008p1423, Carter:2007p297} is inserted into the
  plasma at the far
  end of the depletion, 13~m from the plasma source.  The antenna is
  centered on the depletion and aligned to launch SAWs along the depletion. 
  Each strap of the antenna is driven and controlled independently, giving rise to two
  perpendicularly polarized, co-propagating SAWs.  The frequencies of
  the two waves are offset by a few kHz, and with each frequency near half of
  the ion cyclotron frequency ($f_{\rm ci} = 304$ kHz at 800G).  The
  peak wave magnetic field for SAWs used in this study ranges from
  0.1G to 1G ($1\times 10^{-4} \lesssim \delta B/B \lesssim 1\times
  10^{-3}$). In the absence of the density depletion, the two SAWs
  interact nonlinearly, driving a nonresonant quasimode at their beat
  frequency~\cite{Carter:2006p300}.

\begin{figure}[tbp]
\begin{center}
\includegraphics{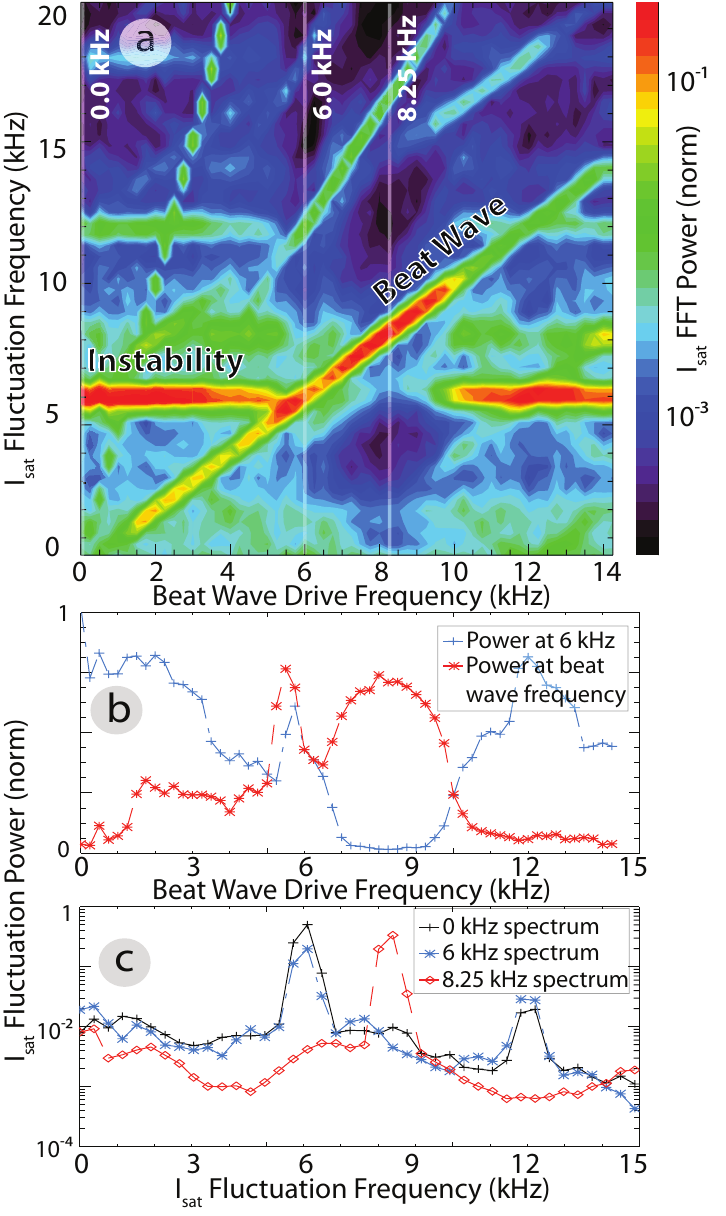}
\caption{[Color]
(Units normalized to power equivalent of maximum RMS fluctuations of 15\%)
(a) Measured $I_{\rm sat}$ fluctuation power as a function of the SAW
  beat frequency (x axis) and FFT frequency bin (y axis).
(b) Power at the spontaneous mode frequency ($f = 6 \pm 0.4$ kHz) and at the SAW beat frequency 
($f = f_{beat} \pm 0.4$ kHz).
(c) $I_{\rm sat}$ fluctuation power spectra at 0.0, 6.0, and 8.25 kHz SAW beat frequency (marked with vertical lines on the 2-D bi-spectral plot above in part a).
}
\label{fig:suppress}
\end{center}
\end{figure}

  To investigate the interaction between the beating SAWs
  and the instability driven by the density depletion, a series of
  experiments were performed in which the beat frequency was set near
  the instability frequency and varied shot-to-shot.
  Fig.~\ref{fig:suppress}(a) shows the measured $I_{\rm sat}$ power
  spectrum near the depletion gradient maximum (at approximately x=2 cm, y=-1 cm (see Fig.~\ref{fig:density} axes) 6 m from the antenna and 7 meters from cathode) as the beat
  frequency is varied over $0 \le f_{\rm beat}  \le 15$ kHz.  

The beat-driven response is observed to be strongest when the beat
frequency is close to the instability frequency (6 kHz), with
peaks centered at 5.5 kHz and 8.25 kHz as shown in Fig.~\ref{fig:suppress}(b).
In this particular dataset, the peak RMS amplitude of the
beat-response is $\delta I_{\rm sat}/I_{\rm sat} = 15\%$.  The
measured beat-wave response is significantly stronger than that
observed in the absence of the depletion, where for the same launched
SAW parameters, a beat-wave amplitude of $\delta I_{\rm sat}/I_{\rm
sat} \lesssim $ 5\% is observed.

The most striking feature of Fig.~\ref{fig:suppress} is the effect
that the SAW beat wave has on the unstable mode
at $f\sim6$ kHz.  As the beat-wave frequency approaches 6 kHz from
below, the spontaneous mode begins to decrease in power and is
observed to downshift in frequency, meeting the beat wave where it
appears to resonate near 5.5 kHz.  As the beat wave is driven to
higher frequency, the spontaneous mode appears completely
absent, as if the growth of the mode is suppressed in the presence
of the driven response.
Fig.~\ref{fig:suppress}(b) illustrates the shift of fluctuation power from the instability to the driven fluctuations
by showing the power
at the unstable mode frequency ($6\pm 0.4$ kHz) and power in the
beat frequency band ($f_{beat} \pm 0.4$ kHz) as a function of
driven beat frequency, essentially a horizontal and diagonal line cut of the data in Fig.~\ref{fig:suppress}(a).
Also of note is the apparent reduction
in broadband fluctuations while the unstable mode is
suppressed.  This is made clearer in Fig.~\ref{fig:suppress}(c),
showing the measured power spectrum in the absence of the SAW beat
wave (both antennas set to the same frequency, black line), and the
spectrum in the presence of a 6 kHz (blue line) and an 8.25 kHz (red line) driven beat wave.  As the beat frequency is
increased further, eventually the instability returns as the
beat-driven response appears to diminish.

\begin{figure}[tbp]
\begin{center}
\includegraphics{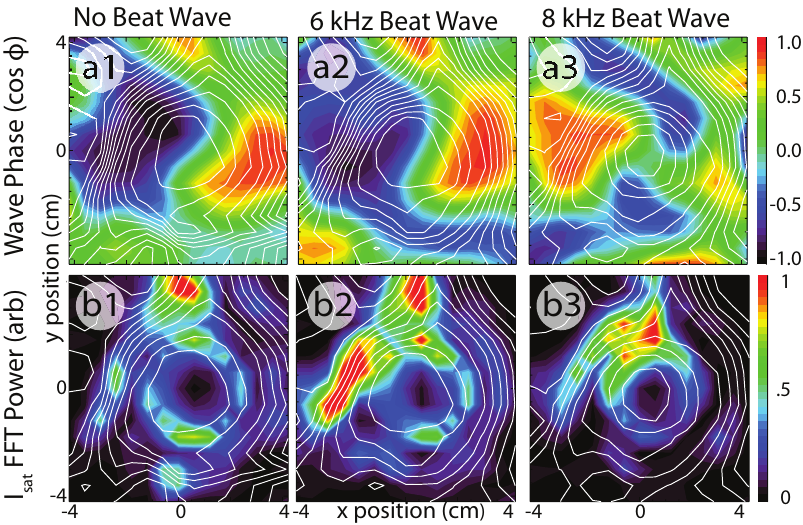}
\caption{[Color] 2D phase (from cross-correlation) and spatial power distribution of
  $I_{\rm sat}$ fluctuations for the spontaneous mode (no
  SAWs) and 6 kHz and 8 kHz beat waves.}
\label{fig:modes}
\end{center}
\end{figure}

The peaks in Fig.~\ref{fig:suppress}(b) suggest that the SAW beat wave
is resonantly driving a linear mode at $f\sim 5$ kHz and $f\sim 8$ kHz
and this conjecture is supported by the spatial structure of the
beat-wave response found using cross-correlation measurements.
Figure~\ref{fig:modes} shows the two-dimensional cross-phase (row a)
and spatial distribution of fluctuation power (row b) for three cases:
the spontaneous mode in the absence of SAWs (column 1), the beat-wave
response when driven at 6 kHz (column 2), and the beat wave response
when driven at 8 kHz (column 3).  Note that for the two beat-wave
cases, the power in the beat-wave response is concentrated on the
density gradient (Figs.~\ref{fig:modes}(b2) and (b3)), similar to the
spontaneous mode (Fig.~\ref{fig:modes}(b1)).  In contrast to this
observation, in the absence of the depletion, the beat-wave
has a much broader spatial distribution consistent with the
pattern of the launched SAWs (not shown here).  The correlation
function of the beat driven response at 6 kHz
(Fig.~\ref{fig:modes}(a2)) is nearly identical to the spontaneous mode
(Fig.~\ref{fig:modes}(a1)) and indicates an $m=-1$ mode structure.
The beat driven response at 6kHz also rotates in the electron
diamagnetic drift direction.  The peak in the beat response at 8kHz
corresponds to an $m=-2$ mode, as shown in Fig.~\ref{fig:modes}(a3). 

\begin{figure}[tbp]
\begin{center}
\includegraphics{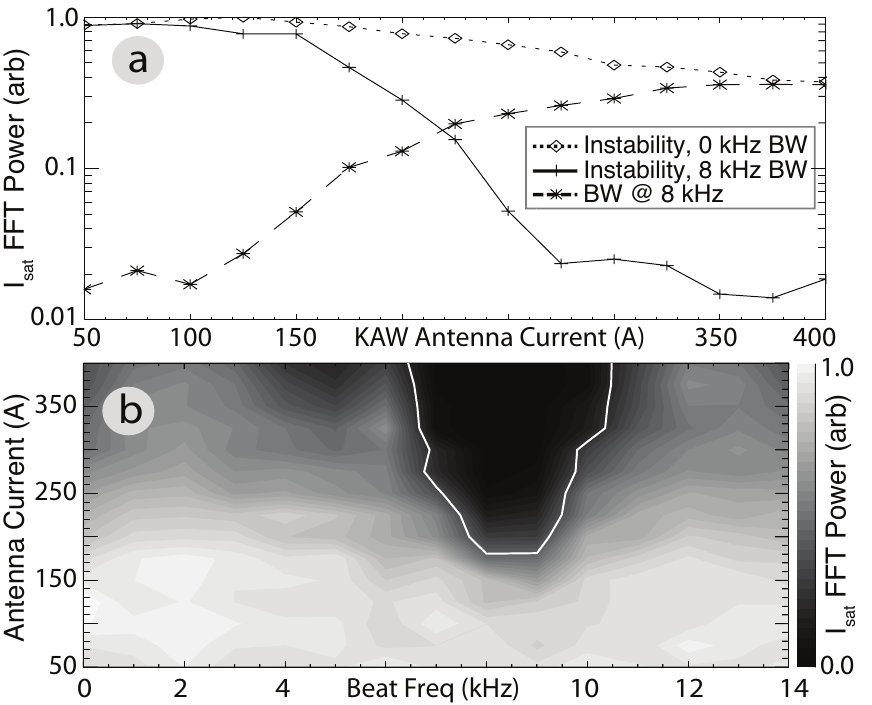}
\caption{(a) Power in the instability band($6.5 \pm 0.4$ kHz, solid line) and
driven beat response band ($8.0 \pm 0.4$ kHz, dashed line) versus SAW drive current for the case of 8.0 kHz beat frequency,
along with a reference instability power ($6.5 \pm 0.4$ kHz, dotted line) 
at zero beat frequency (both antennas at same frequency).
(b) Fluctuation power in the instability band ($6.5 \pm 0.4$ kHz) vs beat-frequency vs antenna current.
The white contour is the 50\% suppressed level at each power.}
\label{fig:suppress_drive}
\end{center}
\end{figure}

The appearance of the control of the instability shown in Fig.~\ref{fig:suppress}
depends strongly on the strength of the beat-wave drive, a parameter
controlled by increasing the amplitude of the launched SAWs.
Fig.~\ref{fig:suppress_drive}(a) shows the variation in the beat-wave
and the instability amplitudes as the SAW antenna
current is increased (for fixed 8kHz beat frequency).  
A threshold is clear for affecting the instability, requiring the beat-wave
amplitude to be on the order of 10\% of the amplitude of the original
unstable mode.  It is also interesting to note that the
beat-wave response saturates as the instability is completely
suppressed.  
Also shown is the instability power as a function of antenna current
in the case of zero beat frequency.  A reduction of the instability
power at high SAW amplitude is observed which can be attributed to
profile modification due to electron heating by the launched SAWs ~\cite{Carter:2007p297}.  
This effect contributes to modification of the instability in
the presence of SAWs, but is less significant than the reduction
observed with non-zero beat frequency.

The width in beat frequency over which control of the instability
is observed also varies with amplitude of the driven beat wave
response.  Fig.~\ref{fig:suppress_drive}(b) shows the FFT power at the
unstable mode frequency as a function of beat wave frequency and
antenna current (which controls the amplitude of the interacting
Alfv\'{e}n waves).
At higher currents, the instability ($m=-1$) begins to be suppressed in favor of the higher frequency mode driven 
by the beat-wave ($m=-2$).  The white contour in
Fig.~\ref{fig:suppress_drive}(b) indicates where the instability power
is at 50\% of the `background' power with the antennas driving a zero
beat frequency Alfv\'en wave.  The broadening of the range over which
the instability is controlled as the beat-wave power is increased is
indicated by this contour.  Similar observations are made in
VINETA~\cite{Brandt:2010p1445}, where the effect was compared to 
``Arnold Tongues'' seen in nonlinearly driven unstable
oscillators~\cite{Schuster:2005p1484,Koepke:1991p1410}.

The exact mechanism by which the beating SAWs couple nonlinearly to
and control the unstable mode is still under investigation.  
Experiments have established that the parallel wavelength of the beat wave is
essential.  Beat waves generated by co-propagating SAWs have very long
parallel wavelength, as long or longer than the length of LAPD at low
frequency~\cite{Carter:2006p300}, similar to expectations for the
parallel wave number of instabilities in the depletion.
Experiments were performed in which counter-propagating SAWs
interacted along the depletion, driving beat waves with much shorter
parallel wavelength.  In this case, much smaller response at the beat
frequency is observed and the control behavior shown in
Fig.~\ref{fig:suppress_drive} is not achieved.

An important question is whether or not transport is affected through
this beat-wave control process.  The unmodified instability has
density-potential crossphase (Figure \ref{fig:density} (e)) that is
inconsistent with fluctuation-driven particle transport.  Past
experiments on filamentary pressure striations with dominant coherent
modes in LAPD have revealed classical transport
levels~\cite{Burke:1998p1416}.  In the presence of the beat-wave in
the experiments reported here, the density-potential crossphase is
unaltered.  Future ex- periments will investigate the interaction
between SAW beat waves and broadband turbulence in LAPD, to see
whether or not transport control can be affected, as has been seen in
mode synchronization experiments on the VINETA
device~\cite{Brandt:2009p1425}.

In summary, a novel approach to control of gradient-
driven instabilities in magnetized laboratory plasmas has
been presented. Beating SAWs have been observed to
nonlinearly interact with and control unstable fluctuations on a
filamentary density depletion in LAPD.  These results provide
motivation for studying the interaction between RF and instabilities
in other magnetized plasmas, including fusion plasmas, e.g. ICRF
beat-wave interaction with microturbulence.  ICRF beat waves have been
successfully used to produce measurable plasma response at the beat
frequency and to excite Alfv\'{e}n eigenmodes in JET and ASDEX
Upgrade~\cite{Fasoli:1996p630, Sassenberg:2010p1485}, motivating
future experiments in which the beat wave frequency is lowered into
the range of drift-wave turbulence.

%\bibliographystyle{prsty}
%\bibliography{suppression_paper}

\end{document}